\documentclass[a4paper,12pt]{article}

\usepackage{amsmath,amssymb,mathtools} 
\usepackage{color}
\usepackage{graphicx}
\usepackage{subfigure}
\usepackage{cite}           
\usepackage{hyperref}            
\usepackage{multirow,makecell}   
\usepackage{textcomp}
\usepackage{wasysym}
\usepackage{setspace}
\usepackage{verbatim}
\usepackage{float}
\usepackage{ulem}
\usepackage[utf8]{inputenc}
\graphicspath{{newFig/}}

\usepackage[text={17.2cm,25.2cm},centering]{geometry}

\numberwithin{equation}{section}

\begin{document}
	\title{\textbf{Strong Coupling Expansion of Gluodynamics on a Lattice under Rotation}}
	\author{Sheng Wang$^{1}$\footnote{shengwang@mails.ccnu.edu.cn},  ~  Jun-Xia Chen$^{1}$\footnote{ chenjunxia@mails.ccnu.edu.cn } ,  ~ Defu Hou$^{1}$\footnote{houdf@mail.ccnu.edu.cn } , ~     Hai-Cang Ren$^{1,2}$\footnote{v-renhc@shanghaitech.edu.cn } }
    \date{}
	
	\maketitle
	
	\vspace{-10mm}
	
	\begin{center}
	{\it
		$^{1}$ Institute of Particle Physics and Key Laboratory of Quark and Lepton Physics (MOS),
		Central China Normal University, Wuhan 430079, China\\ \vspace{1mm}
		
		$^{2}$ School of Physical Science and Technology,  ShanghaiTech University, Shanghai 201210, China\\ \vspace{1mm}			
	}
	\vspace{10mm}
\end{center}

	\vspace{-10mm}
	
	\maketitle
	
	\vspace{-10mm}

	\vspace{10mm}

\begin{abstract}

The analytic strong coupling expansion of the gluodynamics under a rotation with an angular velocity $\omega$ is reported. While the expansion is systematic, free from additional assumptions, the deconfinement temperature determined by the onset of the Polyakov loop expectation value decreases with the angular velocity up to $\omega^2$, opposite to the tendency found in numerical simulations. As a by-product, a simple formula is obtained for the $\omega^2$ coefficient of the deconfinement temperature shift in terms of the latent heat and the discontinuity of the moment of inertia at the transition without rotation. This formula is independent of the strong coupling and may benefit further investigation of the subject.  
	
\end{abstract}

\baselineskip 18pt
\thispagestyle{empty}
\newpage

\tableofcontents

\maketitle

\section{Introduction}\label{sec1}

The experiment of relativistic heavy ion collision revealed many non-perturbative properties of quantum chromodynamics (QCD) under unusual conditions, which stimulated a wide spectrum of theoretical investigations. Among them is the response of the quark-gluon plasma (QGP) to the large angular momentum left behind off-central collisions\cite{STAR:2017ckg, Liang:2004ph} and the behavior of the deconfinement transition temperature under rotation remains an intriguing issue.

The lattice simulation of gluodynamics without dynamic quarks under rotation indicates that the transition temperature increases with increasing angular velocity\cite{Braguta:2021jgn}, while most holographic calculations show opposite dependence \cite{Chen:2020ath,Zhao:2022uxc, Braga:2022yfe,Wang:2024szr} except in Refs.\cite{Chen:2024edy,Chen:2024jet}. For a cylindrical system of radius $R$ rotating about its axis, the transition temperature to the leading order correction by a small linear speed $v=\omega R$ at the boundary is given by
\begin{equation}
T_d=T_d^{(0)}(1+Bv^2),
\label{correction}
\end{equation}
where $T_d^{(0)}$ is the transition temperature without rotation. The lattice simulation gives rise to $B=0.7$\cite{Braguta:2021jgn} for an open boundary condition. All existing researches in this regard are either numerical or relying on the assumption of gauge/gravity duality. A purely analytic approach from field theoretic perspective remains lacking.

An effective action of Polyakov loops as obtained from the strong coupling expansion of the lattice action of the color group $SU(N_c)$ without dynamic quarks \cite{Gross:1983ju} has been applied in the PNJL model to locate the critical ending point \cite{Fukushima:2003fw}. Interestingly, this effective action predicts a second-order deconfinement transition for the $N_c=2$ and a first-order transition for $N_c\geq3$, consistent with simulation results\cite{McLerran:1980pk,Polonyi:1982wz,Kuti:1980gh,Gross:1983ju}. The derivation of the effective action in \cite{Gross:1983ju} can be generalized straightforwardly to include rotation and so we did with the same lattice action employed for the numerical simulation reported in \cite{Braguta:2021jgn}. Our result shows that the transition temperature decreases with the angular velocity, which is in line with most of the holographic calculations. The opposite dependence from the numerical simulation maybe attributed to the limitation of the strong coupling expansion. Even though we were not able to reproduce the simulated angular velocity dependence, our analysis revealed the thermodynamic quantities in the absence of rotation that determine the coefficient $B$ independent of the strong coupling approximation, i.e. 
\begin{equation}
B=-\frac{\Delta I}{2R^2L},
\label{general}
\end{equation}
where $L$ is the latent heat and $\Delta I$ is the discontinuity in the moment of inertia at the transition.
Therefore, we feel it is worthwhile to share our analysis with the interested readers. In addition, we also calculated the string tension and found it enhanced by the rotation.

This paper is organized as follows. In section \ref{sec2}, the general formulation for a thermal equilibrium ensemble is reviewed. In section \ref{sec3}, the lattice action under rotation is presented along with some general formulas related to the effective action of Polyakov loops.
In section \ref{sec4}, the rotation correction of the phase transition point is calculated to the leading order in strong coupling expansion.
Section \ref{sec5} concludes the paper. The relation between different forms for the angular momentum of gluon field is discussed in Appendix A.

\section{A Grand Canonical Ensemble Carrying a Macroscopic Angular Momentum}
\label{sec2}

For a thermal equilibrium ensemble carrying a macroscopic angular momentum component, say $J_z$, the grand partition function is defined by
\begin{equation}
\mathcal{Q}=\rm{Tr}\exp{\left[-\frac{1}{T}(H-\omega J_z)\right]},
\label{partition}
\end{equation}
where $H$ is the Hamiltonian, $T$ is the temperature 
and $\omega$ is the angular velocity. For a Yang-Mills theory without dynamic quarks in the time axial gauge, $A_0^c=0$,

\begin{equation}
H=\frac{1}{2}\int d^3\vec r\left({\vec\Pi}^c{\vec\Pi}^c+{\vec B}^c{\vec B}^c\right),
\end{equation}
where $\vec\Pi^c=-\vec E^c$ is the canonical momentum density with $\vec E^c$ the color electric and $\vec B^c$ the color magnetic field
\begin{equation}
\vec B^c=\vec\nabla\times \vec A^c+\frac{1}{2}
gf^{cab}\vec A^a\times\vec A^b,
\end{equation}
with $\vec A^c$ the gauge potential, $g$ the coupling constant and $a,b,c$ color superscripts. The Gauss law constraint is imposed on the physical states, i.e.
\begin{equation}
\mathcal G|\text{ } 
\rangle=0,
\label{Gauss}
\end{equation}
where 
\begin{equation}
\mathcal{G}=\vec\nabla\cdot\Pi^c+gf^{cab}\vec A^a\cdot\vec\Pi^b,
\end{equation}
serves the generator of a time independent gauge transformation in Hilbert space. The gauge invariant angular momentum operator reads
\begin{equation}
\vec J=\int d^3\vec r[\vec r
\times(\vec B^c\times\vec\Pi^c)]
\label{J_inv},
\end{equation}
which is equivalent to the angular momentum derived from Noether theorem
\begin{equation}
\vec J_{\rm{Noether}}=-\int d^3\vec r\left[\{(\vec r\times\vec\nabla)A_j^c\}\Pi_j^c-\vec A^c\times\vec\Pi^c\right],
\end{equation}
upon dropping a surface term and applying the Gauss law Eq.(\ref{Gauss}). The proof is presented in Appendix A.
As there are no ordering issues between the canonical coordinate $\vec A^c$ and canonical momentum density $\vec\Pi^c$ in the angular momentum term, it is straightforward to derive the Lagrangian for the path-integral through the Legendre transformation\cite{Christ:1980ku}\footnote{According to \cite{Christ:1980ku}, the path-integral formulation requires $\vec A^c$ and $\vec\Pi^c$ to be Weyl-ordered. The difference between the order in the angular momentum density Eq.(\ref{J_inv}) and its Weyl order is merely a constant.}
\begin{equation}
\dot{\vec A}^c(\vec r)=\frac{\delta}{\delta\vec\Pi^c(\vec r)}(H-\omega J_z)=\vec\Pi^c+\omega\vec B^c\times(\hat z\times\vec r),
\end{equation}
and 
\begin{equation}
L=\int d^3\vec r(\vec\Pi^c\cdot\dot{\vec A}^c-H+\omega J_z)=\int d^3\vec r\mathcal{L}.
\end{equation}
We have
\begin{eqnarray}
\mathcal{L} &=& \frac{1}{2}(F_{tx}^cF_{tx}^c+F_{ty}^cF_{ty}^c+F_{tz}^cF_{tz}^c)-\frac{1}{2}(1-\omega^2\rho^2)F_{xy}^cF_{xy}^c\nonumber\\
&-& \frac{1}{2}(1-\omega^2x^2)F_{yz}^cF_{yz}^c-\frac{1}{2}(1-\omega^2y^2)F_{zx}^cF_{zx}^c\nonumber\\
&+& \omega(xF_{xy}^cF_{tx}^c+yF_{xy}^cF_{ty}^c-xF_{yz}^cF_{tz}^c-yF_{zx}^cF_{tz}^c)+\omega^2xyF_{yz}^cF_{zx}^c,
\label{continuum}
\end{eqnarray}
where $\rho^2=x^2+y^2$ in a coordinate system with z-axis the rotation axis. $\mathcal{L}$ corresponds to the Yang-Mills Lagrangian in a global rotating frame with the metric
\begin{equation}
ds^2=-(1-\omega^2\rho^2)dt^2+2\omega(xdy-ydx)dt+dx^2+dy^2+dz^2,
\end{equation}
in the time axial gauge. Restoring nonzero $A_0^c$ in a formal solution of the Gauss law constraint Eq.(\ref{Gauss}), the manifestly gauge invariant action
\begin{equation}
\mathcal{S}=\int Ldt=\int d^4x\mathcal{L},
\label{RealTime}
\end{equation}
together with appropriate gauge fixing terms and Faddeev-Popov determinant define the path integral formulation of the gluodynamics. Its Euclidean version via the continuation
\begin{equation}
t\to -it,\qquad F_{tj}^c\to iF_{tj}^c,\qquad \mathcal{S}\to -i\mathcal{S}_E,
\end{equation}
subject to the periodic boundary condition for the imaginary time interval $0<t<1/T$ gives rise to the partition function Eq.(\ref{partition}). 
The Euclidean version of Eq.(\ref{RealTime}) without gauge fixing is the continuum limit of the lattice action $\mathcal{S}_G$ in the next section. For the strong coupling expansion of the lattice action, we don't have to continuate $\omega$ to an imaginary value.

For the thermodynamics, let us name the free energy corresponding to the partition function Eq.(\ref{partition}) as Gibbs free energy, i.e.
\begin{equation}
\mathcal{F}=-T\ln\mathcal{Q}.
\label{Gibbs}
\end{equation}
The thermal average of the angular momentum reads
\begin{equation}
M=\langle J_z\rangle=-\frac{\partial\mathcal{F}}{\partial\omega}.
\end{equation}
The Helmholtz free energy $F$ follows from the Legendre transformation
\begin{equation}
F=\mathcal{F}+M\omega,
\end{equation}
with
\begin{equation}
\omega=\frac{\partial F}{\partial{M}}.
\end{equation}
For a slow rotation, we have
\begin{equation}
\mathcal{F}=\mathcal{F}^{(0)}-\frac{1}{2}I\omega^2+O(\omega^4),
\end{equation}
and $M=I\omega+O(\omega^3)$, where the superscript $(0)$ signifies the quantities without rotation. Consequently
\begin{equation}
F=F^{(0)}+\frac{1}{2I}M^2+O(M^4),
\end{equation}
with $F^{(0)}=\mathcal{F}^{(0)}$, which gives rise to the meaning of the coefficient $I$ as the moment of inertia.

\section{Lattice action}
\label{sec3}

The simplistic lattice formulation of the thermal equilibrium ensemble is formulated on a hypercubic lattice of size $N_s^3\times N_t$ with lattice spacing $a$ and $N_s\gg N_t$. Periodic boundary condition is imposed in temporal dimension with temperature given by
\begin{equation}
T=\frac{1}{N_ta}.
\end{equation}
The action in the rotating background is\cite{Yamamoto:2013zwa}
\begin{align}
\label{lattice action}
    \mathcal{S}_{G}	&=\frac{2N_{c}}{g^{2}}\sum_{X}[(1-r^{2}\omega^{2})(1-\frac{1}{N_{c}}\text{ReTr}\bar{U}_{xy})+(1-y^{2}\omega^{2})(1-\frac{1}{N_{c}}\text{ReTr}\bar{U}_{xz})\nonumber\\
	&+(1-x^{2}\omega^{2})(1-\frac{1}{N_{c}}\text{ReTr}\bar{U}_{yz})
	+3-\frac{1}{N_{c}}\text{ReTr}(\bar{U}_{x\tau}+\bar{U}_{y\tau}+\bar{U}_{z\tau})\nonumber\\
	&+\frac{1}{N_{c}}\text{ReTr}(iy\omega(\bar{V}_{xy\tau}+\bar{V}_{xz\tau})-ix\omega(\bar{V}_{yx\tau}+\bar{V}_{yz\tau})+xy\omega^{2}\bar{V}_{xzy})].
\end{align}
\begin{figure}[htbp]
    \centering
      \setlength{\abovecaptionskip}{0.5cm}
    \includegraphics[width=1\linewidth]{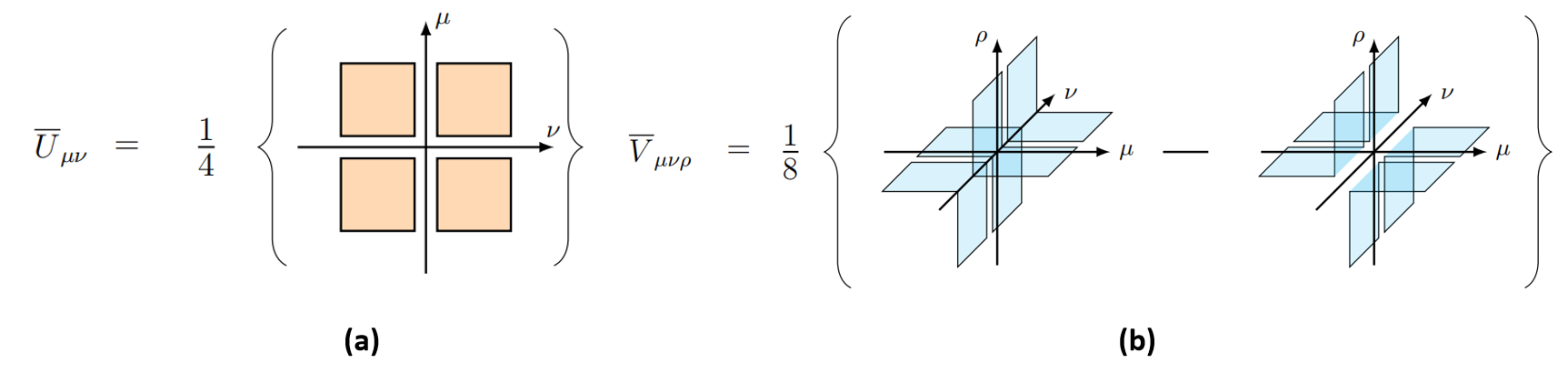}
   \caption{(a) The clover-type average of four plaquettes, (b) the antisymmetric chair-type average of eight chairs from Ref.\cite{Braguta:2021jgn}.}
   \label{plaquettes}
\end{figure}
In the lattice action Eq.(\ref{lattice action}) $X$ under the sum denotes the four dimensional coordinate of a lattice site vector, i.e. $X=(t,\vec r)$ with $t$ the Euclidean time and $\vec {r}$ the spatial coordinate. $\bar{U}_{\mu\nu}$ denotes clover-type average of four plaquettes, $\bar{V}_{\mu\nu\rho}$ is the antisymmetric chair-type average of eight chairs. 
In the continuum limit, we get the following expression for the Euclidean action Eq.(\ref{continuum limit action}).
\begin{align}
    \mathcal{S}_{G}	&=\frac{1}{2g^{2}}\int d^{4}x[(1-r^{2}\omega^{2})F_{xy}^{a}F_{xy}^{a}+(1-y^{2}\omega^{2})F_{xz}^{a}F_{xz}^{a}+(1-x^{2}\omega^{2})F_{yz}^{a}F_{yz}^{a}+F_{x\tau}^{a}F_{x\tau}^{a}\nonumber\\	&+F_{y\tau}^{a}F_{y\tau}^{a}+F_{z\tau}^{a}F_{z\tau}^{a}	-2yi\omega(F_{xy}^{a}F_{y\tau}^{a}+F_{xz}^{q}F_{z\tau}^{a})+2xi\omega(F_{yx}^{q}F_{x\tau}^{a}+F_{yz}^{q}F_{z\tau}^{a})-2xy\omega^{2}F_{xz}^{a}F_{zy}^{a}].
    \label{continuum limit action}
\end{align}

The lattice version of the partition function Eq.(\ref{partition}) reads
\begin{equation}
\mathcal{Q}=\int \prod_{x,\mu}dU_\mu(x)e^{-\mathcal{S}_G[U]}.
\label{partitionL}
\end{equation}
Integrating spatial link variables $U_i(x)$ with $i=x,y,z$ throughout the lattice, the temporal links left over form Polyakov loops
\begin{equation}
W(\vec x)={\rm{tr}}\prod_tU_0(\vec x,t),
\end{equation}
with time ordering imposed on the product, and Eq.(\ref{partitionL}) becomes
\begin{equation}
\mathcal{Q}=\int\prod_{\vec x}dW(\vec x)e^{-\mathcal{S}_{eff}[W,T]}.
\label{gibbs}
\end{equation}
To explore the phase structure of the system, let's introduce a $W$ dependent Gibbs free energy
\begin{align}
    \mathcal{F}[W,T]=T\mathcal{S}_{eff}[W,T],
    \label{mathcal Gibbs}
\end{align}
which carries a global $Z(3)$ symmetry. The Gibbs free energy defined in Eq.(\ref{Gibbs}) at temperature $T$ and angular velocity $\omega$ corresponds to the global minimum of Eq.(\ref{mathcal Gibbs}) with respect to $W(\vec x)$ under saddle point approximation. We have the necessary condition for the global minimum
\begin{equation}
\left(\frac{\partial \mathcal{F}}{\partial W(\vec x)}\right)_{T,\omega}=0,
\label{minimum}
\end{equation}
and $W=0$ is always a solution. A first-order phase transition occurs at the temperature when the minimum corresponding to $W=0$ becomes degenerate with the minimum of the symmetry-breaking solution, $W=\mathcal{W}\neq0$, i.e.
\begin{equation}
\mathcal{F}[0,T_d]=\mathcal{F}[\mathcal{W}, T_d].
\label{first}
\end{equation}
For a rotation with low linear velocity on the boundary, $v=\omega R\ll1$, the free energy to the leading order correction reads
\begin{equation}
\mathcal{F}[W,T]=F_0[W,T]-\frac{1}{2}\omega^2I[W,T]
=F_0[W,T]-\frac{v^2}{2R^2}I[W,T],
\label{free}
\end{equation}
with $F_0[W,T]$ represents the Helmholtz free energy  and $I$ is the moment of inertia. The Gibbs free energy without rotation is the same as Helmholtz free energy. The symmetry breaking solution to Eq.(\ref{minimum}) up to $O(v^2)$ takes the form
\begin{equation}
\mathcal{W}=\mathcal{W}_0+v^2\mathcal{W}_1,
\label{shift}
\end{equation}
where $\mathcal{W}_0(T)$ is the solution to the Eq. (\ref{minimum}) at $v=0$, i.e.
\begin{equation}
\left(\frac{\partial F_0}{\partial \mathcal{W}_0}\right)_T=0,
\label{minimum0}
\end{equation}
and depends only on temperature but $\mathcal{W}_1(T)$ is also coordinate dependent because of the inhomogeneity introduced by rotation. Substituting Eq.(\ref{shift}) into Eq.(\ref{free}), we find that
\begin{equation}
\mathcal{F}[\mathcal{W},T]=F_0[\mathcal{W}_0,T]-\frac{v^2}{2R^2}I[\mathcal{W}_0,T],
\label{free_W}
\end{equation}
and the inhomogeneous correction $\mathcal{W}_1$ does not contribute to this order because of Eq.(\ref{minimum0}).

In the absence of rotation, the condition Eq.(\ref{first}) for the deconfinement phase transition reads
\begin{equation}
F_0[0,T_d^{(0)}]=F_0[\mathcal{W}_0(T_d^{(0)}), T_d^{(0)}].
\label{first0}
\end{equation}
To determine the shift of the transition temperature to the order of $v^2$, we substitute Eq.(\ref{correction}) into Eq.(\ref{first}). We have, to the order of $v^2$, 
\begin{equation}
\mathcal{F}[0,T_d]=F_0[0,T_d^{(0)}]-Bv^2T_d^{(0)}S_0[0,T_d^{(0)}]-\frac{v^2}{2R^2}I[0,T_d^{(0)}],
\label{LHS}
\end{equation}
and
\begin{eqnarray}
\mathcal{F}[\mathcal{W},T_d] &=& F_0[\mathcal{W}_0(T_d^{(0)}+T_d^{(0)}Bv^2),T_d^{(0)}+T_d^{(0)}Bv^2]-\frac{v^2}{2R^2}I[\mathcal{W}_0(T_d^{(0)}+T_d^{(0)}Bv^2),T_d^{(0)}+T_d^{(0)}Bv^2]\nonumber\\
&=& F_0[\mathcal{W}_0(T_d^{(0)}),T_d^{(0)}]
-Bv^2T_d^{(0)}S_0[\mathcal{W}_0(T_d^{(0)}),T_d^{(0)}]-\frac{v^2}{2R^2}I[\mathcal{W}_0(T_d^{(0)}),T_d^{(0)}],
\label{RHS}
\end{eqnarray}
where $S_0=-\frac{\partial F_0}{\partial T}$ is the entropy of the system without rotation and the minimization condition is employed in the expansion Eq.(\ref{RHS}). It follows from Eq.(\ref{first}) and Eq.(\ref{first0}) that
\begin{equation}
    B=-\frac{1}{2R^2}\frac{I[\mathcal{W}_0(T_d^{(0)}),T_d^{(0)}]-I[0,T_d^{(0)}]}{T_d^{(0)}\big{(}S_0[\mathcal{W}_0(T_d^{(0)}),T_d^{(0)}]-S_0[0,T_d^{(0)}]\big{)}}=-\frac{\Delta{I}}{2R^2{L}}.
\end{equation}
In terms of the latent heat
\begin{equation} L=T_d^{(0)}\left(S_0[\mathcal{W}_0(T_d^{(0)}),T_d^{(0)}]-S_0[0,T_d^{(0)}]\right),
\label{latent heat}
\end{equation}
and the discontinuity of the moment of inertia at the transition
\begin{equation}
\Delta I=I[\mathcal{W}_0(T_d^{(0)}),T_d^{(0)}]-I[0,T_d^{(0)}],
\end{equation}
we end up with Eq.(\ref{general}). As the latent heat is always positive because of higher entropy in the deconfinement phase, a lower (higher) moment of inertia in the deconfinement phase implies increasing (decreasing) transition temperature with rotation.  

\section{Strong coupling expansion}\label{sec4}

The deconfinement phase transition of $SU(2)$\cite{Polonyi:1982wz} or $SU(3)$\cite{Gross:1983ju} group without rotation was studied by strong coupling expansion, and the effective potential was obtained analytically. The nature of the transition temperature, i.e. first order for $SU(3)$ and second order for $SU(2)$, has been confirmed by lattice simulation. It would be interesting to see if the angular velocity dependence of the transition temperature revealed in the lattice simulation is also captured in the strong coupling expansion. 

\begin{figure}[htbp]
    \centering
      \setlength{\abovecaptionskip}{0.5cm}
    \includegraphics[width=0.3\linewidth]{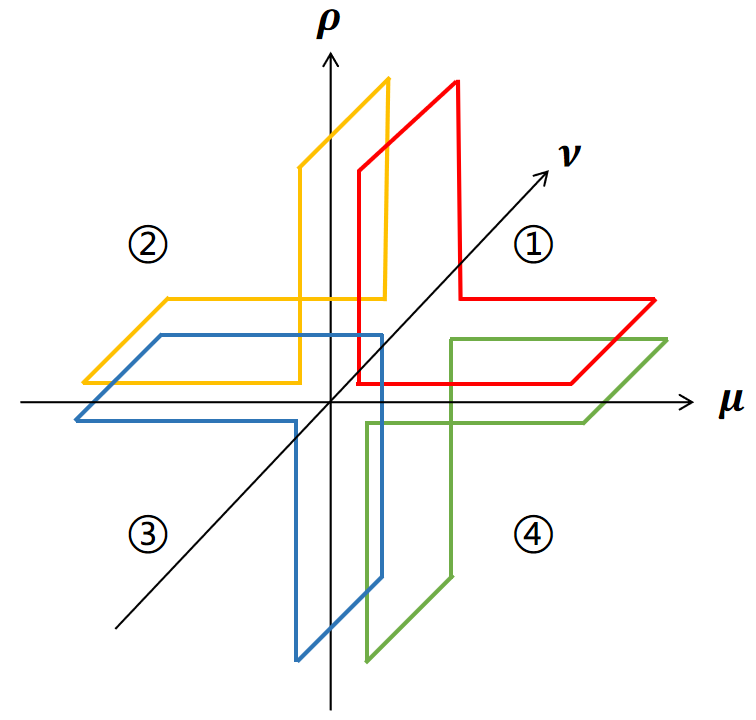}
   \caption{An illustration of the chair-type plaquette $V_{\mu\nu\rho}^s$ with quadrants in $\mu\rho-$plane corresponding different $s$ shown explicitly.}
   \label{quadrant}
\end{figure}

To facilitate the strong coupling expansion in the presence of rotation, we rewrite the lattice action Eq.(\ref{lattice action}) as
\begin{equation}
\mathcal{S}_G=\mathcal{S}_0-\frac{1}{g^2}(E_0+\omega E_1+\omega^2E_2),
\end{equation}
\begin{align}
    E_0=2\text{ReTr}\sum_P(U_{x\tau}+U_{y\tau}+U_{z\tau}+U_{xy}+U_{xz}+U_{yz}),
    \label{E0}
\end{align}

\begin{align}
    E_1=\frac{1}{4}\sum_P\big{\{}\sum_{s=1,2,3,4}(-1)^{s}2\text{ReTr}[i(\bar{y}V_{xy\tau}^s+yV_{xz\tau}^s)-i(\bar{x}V_{yx\tau}^s+xV_{yz\tau}^s)]\big{\}},
    \label{E1}
\end{align}

\begin{align}
    E_2=-2\text{ReTr}\sum_P\big[\bar{r^2}U_{xy}+\bar{y^2}U_{xz}+\bar{x^2}U_{yz}-\frac{1}{4}\sum_{s=1,2,3,4}(-1)^sxyV_{xzy}^s\big],
    \label{E2}
\end{align}
where $\mathcal{S}_0$ is $U$-independent and its dependence on $\omega$ does not impact the phase transition conditions Eq.(\ref{minimum}) and Eq.(\ref{first}).
$\sum_P$ in Eq.(\ref{E0}), Eq.(\ref{E1}) and Eq.(\ref{E2})signifies the summation of the plaquette in each term of the summand throughout the lattice without repetition. For the clover-type plaquette $U_{\mu\nu}$, the coordinate average of the coefficient extends to its four corners, and for a chair-type plaquette $V_{\mu\nu\rho}^s$, the coordinate average of the coefficient extends to the two end points of its hinge in $\nu-$direction. We have $\bar{x}=x$ for $V_{yz\tau}^s$ and $\bar{y}=y$ for $V_{xz\tau}^s$. The superscript $s$ of a chair-type plaquette $V_{\mu\nu\rho}^s$ labels the quadrant in the $\mu\rho$ plane toward which the chair-type plaquette bends the right-hand coordinate system $\mu\nu\rho$. See Fig.\ref{quadrant} for illustration.

\subsection{In the absence of rotation}\label{sec4.1}

\begin{figure}[htbp]
    \centering
      \setlength{\abovecaptionskip}{0.5cm}
    \includegraphics[width=0.75\linewidth]{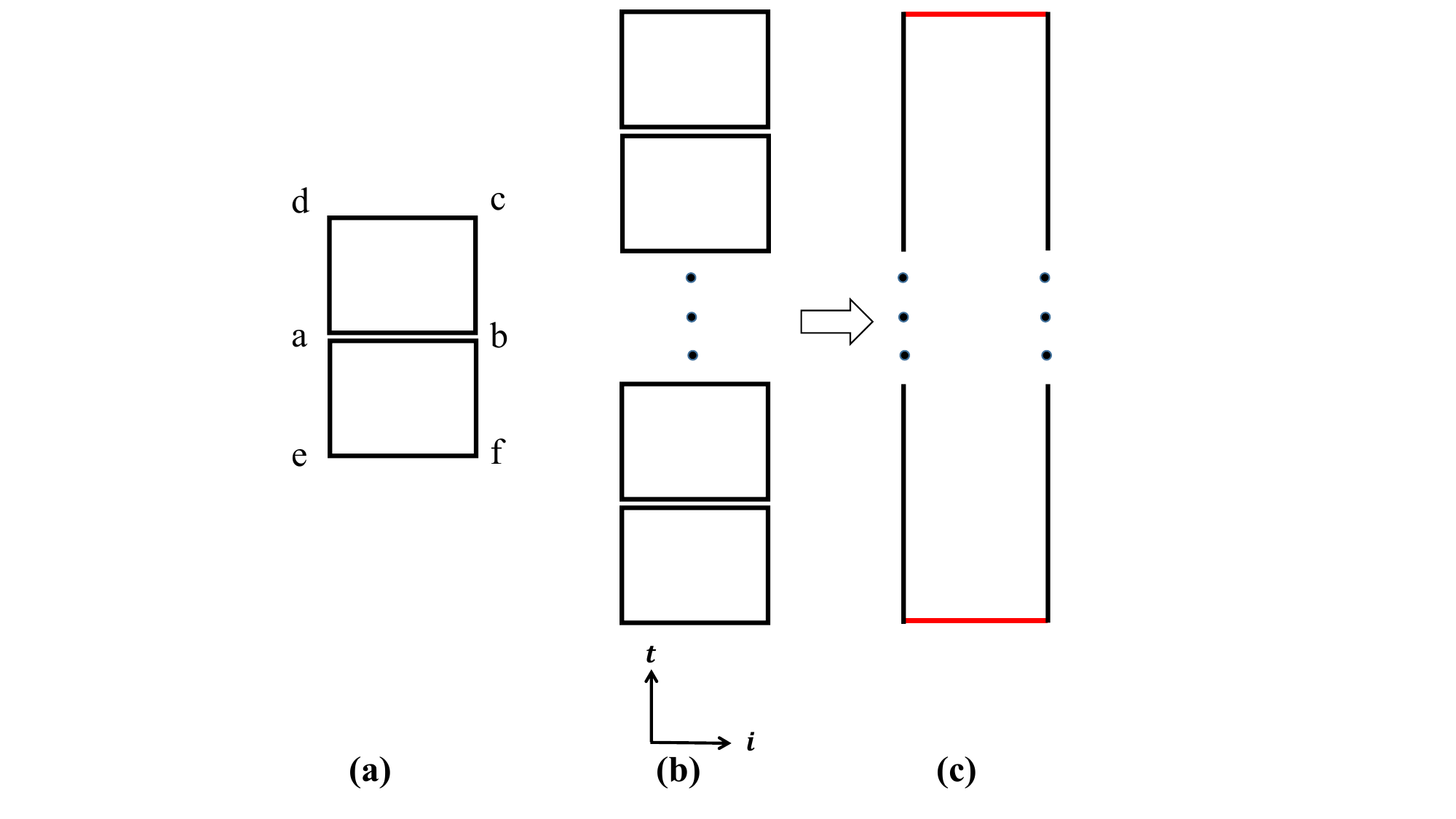}
   \caption{The strong coupling expansion without rotation. (a) An illustration of the integral (\ref{SU(2)}), (b) A ladder diagram to the leading order of strong coupling, (c) The diagram left over after integrating all rungs shared by two plaquettes in (b), resulting in the integrand of (\ref{SU(3)}).} 
   \label{the polyakov loop without rotation}
\end{figure}

To the leading order, the authors of Ref.\cite{Gross:1983ju} found the effective action of $SU(3)$ lattice gauge theory, i.e.
\begin{equation}
\label{eff}
\mathcal{S}_{\rm{eff.}}[W,T]=\frac{1}{2}\lambda\sum_{\vec x, i}|W(\vec x)-W(\vec x+a\hat i)|^2+\sum_{\vec{x}}\mathcal{V}[W(\vec x)],
\end{equation}
where
\begin{equation}
\lambda=2(3g^2)^{-N_t},
\end{equation}
and
\begin{eqnarray}
\mathcal{V}(W,T)&=&-\frac{1}{2}\ln[27-18|W|^2+8\rm{Re}(W^3)-|W|^4]-3\lambda|W|^2\nonumber\\
&=&-\frac{1}{2}\ln[(1+|W|)(3-|W|)^3+8|W|^3(\cos{3\theta}-1)]-3\lambda|W|^2
\label{su(3) effective potential}
\end{eqnarray}
with the logarithmic term stemming from the Haar measure of $SU(3)$ group and $\theta$ the phase of $W$. To derive Eq.(\ref{eff}), the partition function $\mathcal Q$ of Eq.(\ref{partitionL}) is expanded according to the power of $g^{-2}$ as

\begin{equation}
\mathcal{Q}=\int \prod_{x,\mu}dU_\mu(x)e^{-\mathcal{S}_G[U]}
=\text{const.}\int\prod_{x,\mu}dU_\mu(x)\sum_{N=0}^{\infty}\frac{1}{N!g^{2N}}E_0^N,
\label{strong}
\end{equation}
with each term of $E_0^N$ corresponding to a diagram made of plaquettes and the disconnected diagrams are filtered out upon taking logarithm. Then spatial links of each connected diagram are integrated out by orthogonality theorem. To the leading order, $N=N_t$ and the connected diagrams that depend on Polyakov loops and would survive after integrating spatial links consist of $N_t$ plaquettes forming a ladder in temporal dimension as shown in Fig.\ref{the polyakov loop without rotation}(c) with $\vec x$ and $i$ running through the entire spatial lattice and spatial dimensions. Each of the spatial links in Fig.\ref{the polyakov loop without rotation}(b) is covered by a pair of adjacent plaquettes because of the periodic boundary condition. Two key formulas for the integrating spatial links are

\begin{equation}
\int dU_{ab}(\text{tr}U_AU_{ab})(\text{tr}U^\dagger_{ab} U_B)=\frac{1}{3}\text{tr}U_AU_B,
\label{SU(2)}
\end{equation}
and 
\begin{equation}
\int dU{\rm{tr}}UU_CU^\dagger U_D=\frac{1}{3}{\rm{tr}}U_C{\rm{tr}}U_D,
\label{SU(3)}
\end{equation}
where $U_A=U_{bc}U_{cd}U_{da},U_B=U_{ae}U_{ef}U_{fb}$ for Eq.(\ref{SU(2)}). Repeat applications of Eq.(\ref{SU(2)})
to the ladder diagram Fig.\ref{the polyakov loop without rotation}(a) remove the rungs successively until one rung left over as in Fig.\ref{the polyakov loop without rotation}(c). Then the formula (\ref{SU(3)}) is employed with $U_C$ and $U_D$ the products of the $U$ matrices along the two rails of the ladder running opposite directions, giving rise to the product of two $W$'s at adjacent spatial sites. 

Finally, the integration over temporal links is transformed to the integral over the trace of the Polyakov loop operators with the aid of the invariance of the Haar measure and the periodicity of the lattice in temporal dimension. 
As discovered in \cite{Gross:1983ju}, a first order deconfinement phase transition occurs at $\lambda=\lambda_0\approx0.086$, where
\begin{equation}
W=\begin{cases}
\mathcal{W}_0=1+\sqrt{4-\frac{1}{3\lambda}} & \lambda>\lambda_{0},\\
\quad0 & \lambda<\lambda_{0},
\end{cases}
\end{equation}
with $\mathcal{W}_0=1.352$ for $\lambda=\lambda_0^{+}$. In terms of the string tension in strong coupling, extracted from the correlation of two spatially separated Polyakov lines in the confinement phase,
\begin{equation}
\sigma=\frac{1}{a^2}\ln(3g^2),
\label{tension}
\end{equation}
and the lattice spacing $a$, the parameter $\lambda$ is related to temperature as
\begin{equation}
\lambda=2e^{-\frac{\sigma a}{T}}.
\end{equation}
It is straightforward to calculate the latent heat associated with the phase transition and we find that it is positive.
The latent heat of first order deconfinement phase transition is defined in Eq.(\ref{latent heat})
\begin{align}
    L&=T_d^{(0)}(S_0[\mathcal{W}_0(T_d^{(0)}),T_d^{(0)}]-S_0[0,T_d^{(0)}]).
\end{align}
It follows from the Helmholtz free energy $F_0(W,T)=\mathcal{N}T\mathcal{V}(W,T)$ and Eq.(\ref{su(3) effective potential}) for $\mathcal{V}(W,T)$ that the entropy
\begin{equation}
S_0(W,T)=-\frac{\partial F_0}{\partial T}=-\frac{F_0}{T}+6\mathcal{N}\frac{\sigma{a}}{T}e^{-\frac{\sigma{a}}{T}}|W|^2.
\end{equation}
where $\mathcal{N}$ is the number of spatial lattice sites enclosed. Together with  the definition Eq.(\ref{latent heat}) and the phase transition condition Eq.(\ref{first0}), we obtain the latent heat
\begin{equation}
    L=3\lambda_0\mathcal{N}\sigma{a}|\mathcal{W}_0 (T_d^{(0)})|^{2},
\end{equation}
where $\lambda_0=2e^{-\frac{\sigma{a}}{T_d^{(0)}}}$. So we can see that the latent heat is generally positive for first order phase transition. The shift of the phase transition point under rotation will be calculated in the next section.

\subsection{Rotation correction}\label{sec4.2}

\begin{figure}[htbp]
    \centering
      \setlength{\abovecaptionskip}{0.5cm}
    \includegraphics[width=0.76\linewidth]{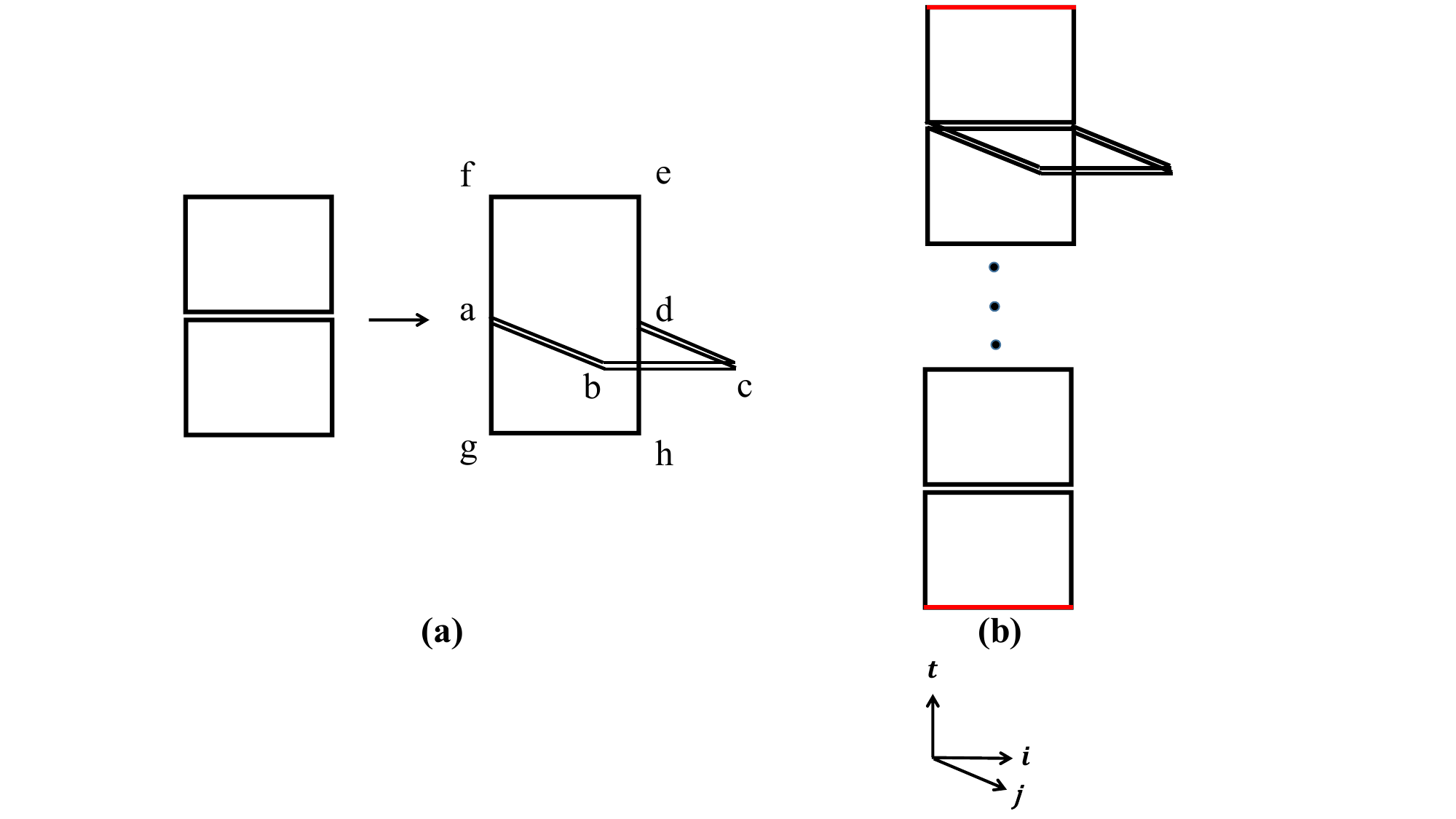}
   \caption{The rotation correction in strong coupling. (a) Left: Replacement of a pair of adjacent square plaquettes by a pair of chair-type plaquettes with the latter corresponding to the integrand of (\ref{chair}), (b) Successive replacement of (a) along the ladder diagram in Fig.\ref{the polyakov loop without rotation}.}
   \label{with rotation}
\end{figure}

In the presence of rotation, the strong coupling expansion Eq.(\ref{strong}) becomes 
\begin{align}
    \mathcal{Q}&=\int \prod_{x,\mu}dU_\mu(x)e^{-\mathcal{S}_G[U]}\nonumber\\
    &=\text{const.}\int\prod_{x,\mu}dU_\mu(x)\sum_{N=0}^{\infty}\frac{1}{N!g^{2N}}(E_{0}+\omega E_{1}+\omega^{2}E_{2})^{N}\nonumber\\
    &=\text{const.}\int\prod_{x,\mu}dU_{\mu}\sum_{N=0}^{\infty}\frac{1}{N!g^{2N}}\big[E_{0}^{N}+C_{N}^1E_{0}^{N-1}(\omega E_{1}+\omega^{2}E_{2})+C_{N}^2\omega^{2}E_{0}^{N-2}E_{1}^{2}+O(\omega^{3})\big],
    \label{strong coupling expansion}
\end{align}
where $C_N^M$'s are binomial coefficients.
For $N=N_t$, the terms inside the brackets in the third line above besides $E_0^{N_t}$ may contribute to the rotation correction of the effective action Eq.(\ref{su(3) effective potential}) up to $\omega^2$ . The single chair-type plaquette from $E_0^{N_t-1}E_1$ or $E_0^{N_t-1}E_2$
will not survive after integrating spatial links since there is no way to replace a square plaquette along the ladder in Fig.\ref{the polyakov loop without rotation}(c) with all spatial links covered by at least two plaquettes and we are left with the term $E_{0}^{N-2}E_{1}^{2}$. The only surviving diagrams are to replace a pair of adjacent square plaquettes in Fig.\ref{the polyakov loop without rotation}(a) with a pair of chair-type plaquettes as shown in Fig.\ref{with rotation}(a), the integration over the three shared spatial links gives rise to

\begin{eqnarray}
&& \int dU_{ab}dU_{bc}dU_{cd}({\rm{tr}}U_AU_{ab}U_{bc}U_{cd})({\rm{tr}}U_{cd}^\dagger U_{bc}^\dagger U_{ab}^\dagger U_B)\nonumber\\
&=&\int dU(\text{tr}U_AU)(\text{tr}U^\dagger U_B)=\frac{1}{3}\text{tr}U_AU_B, 
\label{chair}
\end{eqnarray}

where $U_A=U_{de}U_{ef}U_{fa}, U_B=U_{ag}U_{gh}U_{hd}$. Summing over all possible replacements along the ladder and working out the combinatorial, we end up with the effective action
\begin{equation}
S_{\rm{eff.}}=S_{\rm{eff.}}^{(0)}+S_{\rm{eff.}}^{(1)},
\end{equation}
with $S_{\rm{eff.}}^{(0)}$ given by Eq.(\ref{eff}) and 
\begin{equation}
    S_{\rm{eff.}}^{(1)}=-\frac{\lambda}{4}\omega^2 N_t\sum_{\overrightarrow{r},i}{({\bar{x}^2}+{\bar{y}^2})[W^{\ast}(\overrightarrow{r})W(\overrightarrow{r}+\widehat{i})+W(\overrightarrow{r})W^{\ast}(\overrightarrow{r}+\widehat{i})]},
\end{equation}
where the minus sign stems from the adjacent quadrants which the two chair-type plaquettes in Fig.\ref{with rotation}(a) bend to, and $\bar{x}, \bar{y}$ in front of the bracket refers to the mid point of the link from $\vec r$ to $\vec r+\hat i$.
For a homogeneous configuration, $W(\vec r)=W$ we have
\begin{equation}
S_{\rm{eff.}}^{(1)}=-\frac{1}{4}v^2 \mathcal{N}N_t\lambda\eta W^{\ast}W,
\end{equation}
where we have assumed the same open boundary condition as \cite{Braguta:2021jgn} for a cylindrical volume around the rotation axis with radius $R$\footnote{While the open boundary condition introduces inhomogeneity with respect to the spatial coordinates, we expect that its impact to the transition temperature and order parameter reported in \cite{Gross:1983ju} can be ignored for sufficiently large $R$.}. The coefficient $\eta$ is defined via the average
\begin{equation}
\sum_{\vec r}({\bar{x}^2}+\bar{y}^2)\equiv\frac{1}{2}\eta\mathcal{N}R^2,
\end{equation}
and is $\eta\simeq 1$ for sufficiently large $R/a$ ($\eta=1$ in continuum limit).

It follows from Eqs.(\ref{free_W}) and (\ref{first0})  that the discontinuity in the moment of inertia is
\begin{equation}
\Delta I=I[\mathcal{W}_0(T_d^{(0)}),T_d^{(0)}]-I[0,T_d^{(0)}]=\frac{1}{2}\mathcal{N}N_t\lambda_{0}\eta R^2|\mathcal{W}_0(T_d^{(0)})|^2.
\end{equation}
Together with the latent heat calculated in the previous subsection and the formula (\ref{general}), we obtain that
\begin{equation}
B=-\frac{1}{12\sigma a^2}\approx-0.02235,
\label{B}
\end{equation}
with reference to \cite{Braguta:2021jgn}, we set $\sigma=(440\text{MeV})^2, a^{-1}=228\text{MeV}$.

The same methodology can be applied for calculating the rotation correction to the string tension. The inhomogeneity embedded in the global rotation metric complicates the result. As a simple example, let us consider a pair of Polyakov lines located at coordinates $\vec x_1=(x,y,z)$ and $\vec x_2=(x,y,z+L)$ with $L=Ma$ and calculate their correlation in the confinement phase
\begin{equation}
e^{-\beta U(L)}=\frac{1}{\mathcal{Q}}\int \prod_{x,\mu}dU_\mu(x)e^{-\mathcal{S}_G[U]}W(\vec x_1)W^*(\vec x_2),
\end{equation}
\begin{figure}[htbp]
    \centering
      \setlength{\abovecaptionskip}{0.01cm}
    \includegraphics[width=0.4\linewidth]{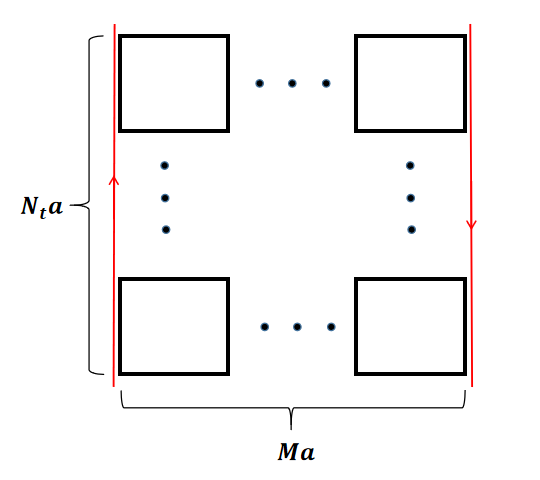}
   \caption{A pair of Polyakov lines apart from $L$ in the confinement phase. The diagram consists of $N_{t}M$ square plaquettes covering the minimum area.}
   \label{string tension}
\end{figure}
with $U(L)$ the potential energy of two heavy quarks located at $\vec x_1$ and $\vec x_2$. It is straightforward to show that the leading contribution to the string tension in strong coupling resides in 
\begin{align}
e^{-\beta U(L)} &= \frac{1}{(N_tM)!}\left(\frac{1}{g^2}\right)^{N_tM}\int \prod_{x,\mu}dU_\mu(x)W(\vec x_1)W^*(\vec x_2)(E_{0}+\omega E_{1}+\omega^{2}E_{2})^{N_tM}\nonumber\\
&= \frac{1}{(N_tM)!}\left(\frac{1}{g^2}\right)^{N_tM}\int \prod_{x,\mu}dU_\mu(x)W(\vec x_1)W^*(\vec x_2)(E_{0}^{N_tM}+C_{N_tM}^2\omega^2E_{0}^{N_tM-2}E_{1}^2).
\end{align}

The connected diagram originated from $E_{0}^{N_tM}$ consists of $N_tM$ square plaquettes covering the minimum area between $W(\vec x_1)$ and $W^*(\vec x_2)$ as shown in Fig.\ref{string tension} and gives rise to Eq.(\ref{tension}). The $E_{0}^{N_tM-2}E_{1}^2$ term replaces a pair of square plaquettes sharing a common link in z-direction with a pair of chair-type plaquettes. It follows from similar logistics for the rotation correction of the effective action that
\begin{equation}
e^{-\beta U(L)}=\left(\frac{1}{3g^2}\right)^{N_tM}(1+N_tM\omega^2d^2),
\end{equation}
with $d$ the distance to the rotation axis and we end up the the string tension up to $\omega^2$
\begin{equation}
\sigma=\frac{\ln(3g^2)+\omega^2d^2}{a^2}.
\end{equation}

\section{Concluding remarks}\label{sec5}

In summary, we started with a grand canonical ensemble of gluon system with a macroscopic angular momentum in Hamiltonian formulation and transformed it into the Matsubara form of path integral. We found that the underlying Lagrangian besides the gauge fixing term is exactly the Lagrangian in a global rotation frame. Within the mean-field framework for the effective action of the Polyakov loop, we related the shift of the deconfinement temperature for a low angular velocity to the latent heat and the jump of the moment of inertia at the transition. 

Upon the strong coupling expansion of the lattice action Eq.(\ref{lattice action}) following the methodology of \cite{Gross:1983ju,Polonyi:1982wz}, we found that the deconfinement temperature defined at the onset of a nonzero thermal average of the Polyakov loop decreases with the angular velocity\footnote{Upon completion of this work, we learned that Dr. K. Fukushima’s team has been pursuing a parallel research.}, opposite to the tendency revealed in lattice simulation but in line with most of holographic results \cite{Chen:2020ath,Zhao:2022uxc, Braga:2022yfe,Wang:2024szr}. This discrepancy maybe attributed to the accuracy of the strong coupling. In view of the small magnitude of the coefficient $B$ in Eq.(\ref{B}), the coefficient of the $\omega^2$ term to the leading order in $g^{-2}$ may not be robust against higher order corrections when extrapolated to the deconfinement transition point. We also calculated the string tension to the order of $\omega^2$ for a pair heavy quark and antiquark separated in parallel to the rotation axis and found it enhanced by the angular velocity. 

The continuum limit of a lattice gauge theory is approached for the coupling lying within the critical window but the strong coupling region is outside of it. On the other hand, the detailed expression of the lattice actions that share the same critical behavior are not unique. The discrepancy noted in this work may be attributed to the simplicity of the Wilson's lattice action.   

Though the purely analytic strong coupling expansion of the lattice formulation is unable to explain the  tendency of deconfinement temperature with respect to the angular velocity discovered in numerical simulation, we would like to share our analysis which may shed light for further investigation. 

\section*{Acknowledgements}

This work is  supported partly by the National Key Research and Development Program of China under Contract No. 2022YFA1604900, and by the National Natural Science Foundation of China (NSFC) under Grant No.12435009 and No. 12275104.

\appendix
\section{Consistency of angular momentum}
\label{appendix A}
In this appendix, we prove that  the gauge
invariant angular momentum operator is equivalent to the angular momentum derived from Noether theorem.

The gauge
invariant angular momentum operator reads
\begin{align}
\vec J&=\int d^3\vec r[\vec r
\times(\vec B^c\times\vec\Pi^c)]\nonumber\\
&=-\int d^3\vec{r}[(\vec{r}\cdot\vec{B}^c)\vec{\Pi}^c-\vec{B}^c(\vec{r}\cdot\vec{\Pi}^c)].
\end{align}
Written in component form, we have
\begin{align}
J_{i}&=-\int d^{3}\vec{r}(r_{l}B_{l}^{c}\Pi_{i}^{c}-r_{j}\Pi_{j}^{c}B_{i}^{c})\nonumber\nonumber\\
&=-\int d^{3}\vec{r}[x_{l}\epsilon_{lkj}(\partial_{k}A_{j}^{c}+\frac{1}{2}gf^{cab}A_{k}^{a}A_{j}^{b})\Pi_{i}^{c}-\epsilon_{ikl}(\partial_{k}A_{l}^{c}+\frac{1}{2}gf^{cab}A_{k}^{a}A_{l}^{b})x_{j}\Pi_{j}^{c}].
\label{J_i}
\end{align}
With the aid of the Schouten identity
\begin{align}
\epsilon_{lkj}\Pi_{i}^c=\epsilon_{lki}\Pi_{j}^c+\epsilon_{lij}\Pi_{k}^c+\epsilon_{ikj}\Pi_{l}^c,
\label{schouten identity}
\end{align}
substitute Eq.(\ref{schouten identity}) into Eq.(\ref{J_i}).Then we rewrite $J_i$
\begin{align}
    J_i&=-\int d^{3}\vec{r}[\epsilon_{lki}x_{l}(\partial_{k}A_{j}^{c}+\frac{1}{2}gf^{cab}A_{k}^{a}A_{j}^{b})\Pi_{j}^{c}+\epsilon_{lij}x_{l}(\partial_{k}A_{j}^{c}+\frac{1}{2}gf^{cab}A_{k}^{a}A_{j}^{b})\Pi_{k}^{c}\nonumber\\
    &\quad\quad +\epsilon_{ikj}x_{l}(\partial_{k}A_{j}^{c}+\frac{1}{2}gf^{cab}A_{k}^{a}A_{j}^{b})\Pi_{l}^{c}-\epsilon_{ikl}(\partial_{k}A_{l}^{c}+\frac{1}{2}gf^{cab}A_{k}^{a}A_{l}^{b})x_{j}\Pi_{j}^{c}]\nonumber\nonumber\\
    &=-\int d^{3}\vec{r}[\epsilon_{lki}x_{l}\partial_{k}A_{j}^{c}\Pi_{j}^{c}+\epsilon_{lij}x_{l}\partial_{k}A_{j}^{c}\Pi_{k}^{c}+\frac{1}{2}g(\epsilon_{lki}x_{l}f^{cab}A_{k}^{a}A_{j}^{b}\Pi_{j}^{c}+\epsilon_{lij}x_{l}f^{cab}A_{k}^{a}A_{j}^{b}\Pi_{k}^{c})]\nonumber\\
    &=-\int d^{3}\vec{r}[\epsilon_{lki}x_{l}\partial_{k}A_{j}^{c}\Pi_{j}^{c}-\epsilon_{lij}x_{l}A_{j}^{c}(\partial_{k}\Pi_{k}^{c})-\epsilon_{lij}A_{j}^{c}\Pi_{l}^{c}+\frac{1}{2}g\epsilon_{lki}x_{l}A_{k}^{a}f^{cab}A_{j}^{b}\Pi_{j}^{c}-\frac{1}{2}g\epsilon_{lji}x_{l}A_{j}^{b}f^{cab}A_{k}^{a}\Pi_{k}^{c}]\nonumber\\
    &=-\int d^3\vec{r}[(\vec{r}\times\vec{\nabla})_{i}A_{j}^{c}\Pi_{j}^{c}-(\vec{A^{c}}\times\vec{\Pi^{c}})_{i}+(\vec{r}\times\vec{A^{c}})_{i}(\vec{\nabla}\cdot\vec{\Pi^{c}}+gf^{cab}\vec{A^{a}}\cdot\vec{\Pi^{b}})],
\end{align}
where we have conducted an integration by part for the second term inside the bracket of the third line above and dropped the surface term. It follows that
\begin{equation}
    \vec{J}=\vec{J}_{\text{Noether}}-\int d^3\vec{r}[(\vec{r}\times\vec{A^c})\mathcal{G}]
    \label{J},
\end{equation}
with
\begin{equation}
\vec{J}_{\text{Noether}}=\int d^3\vec{r}\lbrace\vec{A^{c}}\times\vec{\Pi^{c}}-[(\vec{r}\times\ \vec{\nabla})A_{j}^{c}]\Pi_{j}^{c}\rbrace,
\end{equation}
the angular momentum derived from Noether theorem. According to the Gauss theorem given by Eq.(\ref{Gauss}), the surface term in the second term of Eq.(\ref{J}) can be eliminated. Therefore, we have fully proved that the gauge invariant angular momentum operator is equivalent
to the angular momentum derived from Noether theorem.

	
	
	
\bibliographystyle{utphys}
\bibliography{ref}

\providecommand{\href}[2]{#2}\begingroup\raggedright\begin{thebibliography}{10}

\bibitem{STAR:2017ckg}
{\bfseries STAR} Collaboration, L.~Adamczyk {\em et~al.}, ``{Global $\Lambda$
  hyperon polarization in nuclear collisions: evidence for the most vortical
  fluid},'' \href{http://dx.doi.org/10.1038/nature23004}{{\em Nature}
  {\bfseries 548} (2017) 62--65},
  \href{http://arxiv.org/abs/1701.06657}{{\ttfamily arXiv:1701.06657
  [nucl-ex]}}.

\bibitem{Liang:2004ph}
Z.-T. Liang and X.-N. Wang, ``{Globally polarized quark-gluon plasma in
  non-central A+A collisions},''
  \href{http://dx.doi.org/10.1103/PhysRevLett.94.102301}{{\em Phys. Rev. Lett.}
  {\bfseries 94} (2005) 102301},
  \href{http://arxiv.org/abs/nucl-th/0410079}{{\ttfamily
  arXiv:nucl-th/0410079}}.

\bibitem{Braguta:2021jgn}
V.~V. Braguta, A.~Y. Kotov, D.~D. Kuznedelev, and A.~A. Roenko, ``{Influence of
  relativistic rotation on the confinement-deconfinement transition in
  gluodynamics},'' \href{http://dx.doi.org/10.1103/PhysRevD.103.094515}{{\em
  Phys. Rev. D} {\bfseries 103} no.~9, (2021) 094515},
  \href{http://arxiv.org/abs/2102.05084}{{\ttfamily arXiv:2102.05084
  [hep-lat]}}.

\bibitem{Chen:2020ath}
X.~Chen, L.~Zhang, D.~Li, D.~Hou, and M.~Huang, ``{Gluodynamics and
  deconfinement phase transition under rotation from holography},''
  \href{http://dx.doi.org/10.1007/JHEP07(2021)132}{{\em JHEP} {\bfseries 07}
  (2021) 132}, \href{http://arxiv.org/abs/2010.14478}{{\ttfamily
  arXiv:2010.14478 [hep-ph]}}.

\bibitem{Zhao:2022uxc}
Y.-Q. Zhao, S.~He, D.~Hou, L.~Li, and Z.~Li, ``{Phase diagram of holographic
  thermal dense QCD matter with rotation},''
  \href{http://dx.doi.org/10.1007/JHEP04(2023)115}{{\em JHEP} {\bfseries 04}
  (2023) 115}, \href{http://arxiv.org/abs/2212.14662}{{\ttfamily
  arXiv:2212.14662 [hep-ph]}}.

\bibitem{Braga:2022yfe}
N.~R.~F. Braga, L.~F. Faulhaber, and O.~C. Junqueira,
  ``{Confinement-deconfinement temperature for a rotating quark-gluon
  plasma},'' \href{http://dx.doi.org/10.1103/PhysRevD.105.106003}{{\em Phys.
  Rev. D} {\bfseries 105} no.~10, (2022) 106003},
  \href{http://arxiv.org/abs/2201.05581}{{\ttfamily arXiv:2201.05581
  [hep-th]}}.

\bibitem{Wang:2024szr}
J.-H. Wang and S.-Q. Feng, ``{Rotation effect on the deconfinement phase
  transition in holographic QCD},''
  \href{http://dx.doi.org/10.1103/PhysRevD.109.066019}{{\em Phys. Rev. D}
  {\bfseries 109} no.~6, (2024) 066019},
  \href{http://arxiv.org/abs/2403.01814}{{\ttfamily arXiv:2403.01814
  [hep-ph]}}.

\bibitem{Chen:2024edy}
J.-X. Chen, S.~Wang, D.~Hou, and H.-C. Ren, ``{String tension and Polyakov loop
  in a rotating background},''
  \href{http://dx.doi.org/10.1103/PhysRevD.111.026020}{{\em Phys. Rev. D}
  {\bfseries 111} no.~2, (2025) 026020},
  \href{http://arxiv.org/abs/2410.04763}{{\ttfamily arXiv:2410.04763
  [hep-ph]}}.

\bibitem{Chen:2024jet}
Y.~Chen, X.~Chen, D.~Li, and M.~Huang, ``{Deconfinement and chiral restoration
  phase transition under rotation from holography in an anisotropic
  gravitational background},''
  \href{http://dx.doi.org/10.1103/PhysRevD.111.046006}{{\em Phys. Rev. D}
  {\bfseries 111} no.~4, (2025) 046006},
  \href{http://arxiv.org/abs/2405.06386}{{\ttfamily arXiv:2405.06386
  [hep-ph]}}.

\bibitem{Gross:1983ju}
M.~Gross, J.~Bartholomew, and D.~Hochberg, ``{SU(N) Deconfinement Transition
  and the N State Clock Model},''. Report No. EFI- 83-35-CHICAGO, 1983.

\bibitem{Fukushima:2003fw}
K.~Fukushima, ``{Chiral effective model with the Polyakov loop},''
  \href{http://dx.doi.org/10.1016/j.physletb.2004.04.027}{{\em Phys. Lett. B}
  {\bfseries 591} (2004) 277--284},
  \href{http://arxiv.org/abs/hep-ph/0310121}{{\ttfamily arXiv:hep-ph/0310121}}.

\bibitem{McLerran:1980pk}
L.~D. McLerran and B.~Svetitsky, ``{A Monte Carlo Study of SU(2) Yang-Mills
  Theory at Finite Temperature},''
  \href{http://dx.doi.org/10.1016/0370-2693(81)90986-2}{{\em Phys. Lett. B}
  {\bfseries 98} (1981) 195}.

\bibitem{Polonyi:1982wz}
J.~Polonyi and K.~Szlachanyi, ``{Phase Transition from Strong Coupling
  Expansion},'' \href{http://dx.doi.org/10.1016/0370-2693(82)91280-1}{{\em
  Phys. Lett. B} {\bfseries 110} (1982) 395--398}.

\bibitem{Kuti:1980gh}
J.~Kuti, J.~Polonyi, and K.~Szlachanyi, ``{Monte Carlo Study of SU(2) Gauge
  Theory at Finite Temperature},''
  \href{http://dx.doi.org/10.1016/0370-2693(81)90987-4}{{\em Phys. Lett. B}
  {\bfseries 98} (1981) 199}.

\bibitem{Christ:1980ku}
N.~H. Christ and T.~D. Lee, ``{Operator Ordering and Feynman Rules in Gauge
  Theories},'' \href{http://dx.doi.org/10.1103/PhysRevD.22.939}{{\em Phys. Rev.
  D} {\bfseries 22} (1980) 939}.

\bibitem{Yamamoto:2013zwa}
A.~Yamamoto and Y.~Hirono, ``{Lattice QCD in rotating frames},''
  \href{http://dx.doi.org/10.1103/PhysRevLett.111.081601}{{\em Phys. Rev.
  Lett.} {\bfseries 111} (2013) 081601},
  \href{http://arxiv.org/abs/1303.6292}{{\ttfamily arXiv:1303.6292 [hep-lat]}}.

\end{thebibliography}\endgroup

\end{document}